\documentclass{elsart3}
\usepackage{graphicx,amssymb}
\journal{Nuclear Instruments and Methods in Physics Research B}
\begin{document}
\begin{frontmatter}

\title{Plans for laser spectroscopy of trapped cold hydrogen-like HCI}
\author{D.F.A. Winters\corauthref{cor}},
\corauth[cor]{Corresponding author.}
\ead{d.winters@imperial.ac.uk}
\author{A.M. Abdulla, J.R. Castrej\'on Pita, A. de Lange}, 
\author{D.M. Segal, R.C. Thompson}
\address{Blackett Laboratory, Imperial College, Prince Consort Road, London SW7 2BW, United Kingdom}

\begin{abstract}
Laser spectroscopy studies are being prepared to measure the $1s$ ground state hyperfine splitting in trapped cold highly charged ions. The purpose of such experiments is to test quantum electrodynamics in the strong electric field regime. These experiments form part of the HITRAP project at GSI. A brief review of the planned experiments is presented.
\end{abstract}

\end{frontmatter}

\section{Introduction}
An accurate measurement of the hyperfine splitting (HFS) of the $1s$ ground state of hydrogen-like highly charged ions (HCI) is a good test of quantum electrodynamics (QED) in the limit of strong electric fields ($10^{15}$~V/cm \cite{bei00}). Such strong fields cannot be produced using conventional laboratory techniques, but naturally exist close to the stripped nuclei of heavy elements like Pb, Bi or U. By detecting the fluorescence from the laser-excited upper hyperfine state of such a trapped and cold ion, a high-precision measurement of the hyperfine splitting can be made.

HCI with extremely high charge states ({\it e.g.} Pb$^{81+}$ or U$^{92+}$) and relativistic energies (400~MeV/u) are created in the heavy ion facility (SIS) at GSI in Darmstadt. These HCI are injected into the Experimental Storage Ring (ESR) and can be decelerated (down to 4~MeV/u) forming 1~$\mu$s long bunches containing about $10^5$ ions, arriving every 10 seconds. In the HITRAP project \cite{qui01,tdr03} these bunches will be extracted from the ESR, decelerated by linear (IH-LINAC) and radiofrequency (RFQ) stages, trapped and cooled in a Penning trap (cooler trap), and made available for experiments.

In the ESR at GSI, previous measurements of the HFS were made on bunches of relativistic HCI such as $^{209}$Bi$^{82+}$ \cite{kla94} and $^{207}$Pb$^{81+}$ \cite{see98}. HFS measurements were also made at the SuperEBIT on $^{165}$Ho$^{66+}$ \cite{cre96}, $^{185,187}$Re$^{74+}$ \cite{cre98} and $^{203,205}$Tl$^{80+}$ \cite{bei01}. The resolution obtained in the above experiments is mainly limited by the Doppler effect. A measurement of the $1s$ ground state HFS of trapped cold HCI using laser spectroscopy should be even more accurate due to a cryogenic UHV environment, high ion cloud density, the absence of a large Doppler shift and virtually unlimited measurement time.

Laser spectroscopy offers the possibility of high-accuracy measurements of transition wavelengths in the visible region \cite{tho85}. In HCI, electronic transitions are generally in the far UV or X-ray regions of the spectrum. However, since the $1s$ groundstate HFS scales with the atomic number $Z$ as $Z^3$, the ground state HFS of hydrogen-like HCI can move into the visible spectrum for $Z>70$ \cite{bei00,sha94}. The lifetime of this transition falls as $Z^{-9}$ and is of the order of milliseconds for $Z>70$. A measurement of this transition wavelength gives information on the QED corrections to the HFS or on the spatial distribution of the nuclear magnetisation (Bohr-Weisskopf effect), which is affected by core polarisation and is not really well understood \cite{bei00}. Its measurement thus allows for critical tests of nuclear models. From a comparison of measurements of the HFS of hydrogen-like and lithium-like HCI the nuclear effects can be eliminated so that an accurate measurement of the QED effects can be made \cite{tom98,sun98}.

There are several candidate systems that can be studied at HITRAP, including radioactive isotopes. An interesting first challenge would be to measure the HFS of the $1s$ ground state $M1$ transition of $^{207}$Pb$^{81+}$ ions ($\lambda=1020$~nm \cite{see98}) by means of laser spectroscopy. In order to reach the necessary accuracy, the HCI will be trapped in a cryogenic UHV environment. Once trapped, the ions can be easily stored for long times, therefore the lifetime $\tau=50$~ms \cite{see98} of the upper hyperfine state is not a problem. Electron capture (neutralisation) by collisions is strongly reduced by operating the trap at cryogenic temperatures (4~K) under UHV conditions (below $10^{-14}$~mbar).

\begin{figure}[!t]
\begin{center}
\centering
\includegraphics[width=7.5cm]{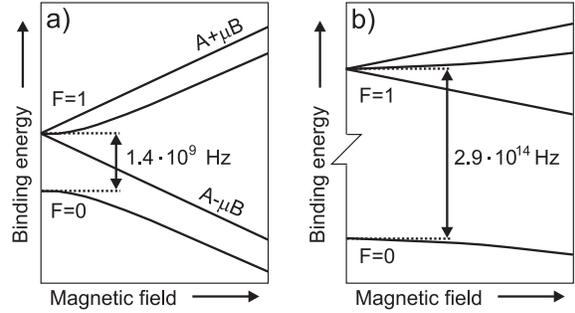}
\caption{Zeeman and hyperfine splittings of the $1s$ ground states of $I=1/2$ nuclei plotted versus magnetic field: a) hydrogen atom $^1$H$^0$, b) highly charged lead ion $^{207}$Pb$^{81+}$. (The plots are drawn with different scales.)}
\label{fig1}
\end{center}
\end{figure}

\section{Zeeman versus hyperfine splitting}
The high precision measurements of the HFS will be performed on HCI in a Penning trap, but the strong magnetic field will shift and split the hyperfine levels and thus the natural HFS. This `Zeeman effect' \cite{zee96} is schematically illustrated in figure 1a) for the $1s$ ground state HFS in hydrogen, which has nuclear spin $I=1/2$. For hydrogen the HFS can be easily calculated using the famous Breit-Rabi formula \cite{bre31}. The zero-field splitting $A$ is modified by the interaction between the magnetic moment $\mu$ of the ion and the magnetic field strength $B$. In the simple hydrogen case at a field of 1~T, the Zeeman splitting of about $2.8 \cdot 10^{10}$~Hz is more than one order of magnitude larger than the HFS in hydrogen ($1.4 \cdot 10^9$~Hz or 21~cm). However, for HCI the HFS dominates the Zeeman splitting by several orders of magnitude. This is indicated in figure 1b) for the $1s$ ground state HFS of $^{207}$Pb$^{81+}$, which also has nuclear spin $I=1/2$.

\begin{figure}[!b]
\begin{center}
\centering
\includegraphics[width=7cm]{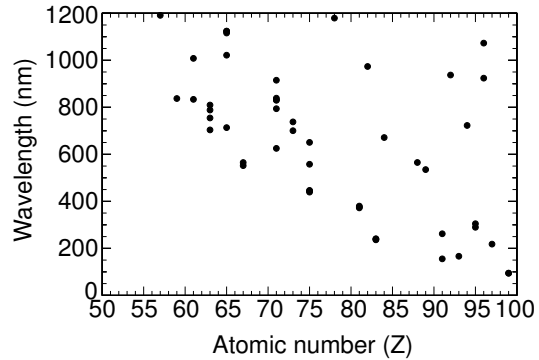}
\caption{Wavelengths of the $1s$ ground state hyperfine splittings in the visible spectrum for atomic number $Z$ ranging from 50 to 100.}
\label{fig2}
\end{center}
\end{figure}

In figure 2 the calculated wavelengths of the $1s$ ground state hyperfine transition (neglecting QED and the Bohr-Weisskopf effect) are plotted for HCI with atomic number $Z$ ranging from 50 to 100. Most of these heavier elements have their HFS in the visible region of the spectrum and are thus easily excited by conventional tunable laser systems.

\section{The spectroscopy trap}
The HCI will be extracted from the HITRAP cooler trap in a long narrow bunch with an average kinetic energy of a few eV and containing about $10^5$ ions. The radial energy spread will be reduced so as to maintain a parallel beam \cite{tdr03}. The HCI will then be loaded into the spectroscopy trap, which is a cylindrical open-endcap Penning trap \cite{gab89} with compensation electrodes to create a nearly perfect quadrupole potential at the trap centre. Figure 3 schematically shows the expected layout of the electrodes, the tank circuit for resistive cooling \cite{win75,ver04} and the trap loading scheme: the HCI enter from the right, are reflected by the left capture electrode, enclosed by the right one, localised near the trap centre and finally cooled and compressed.

\begin{figure}[!b]
\begin{center}
\centering
\includegraphics[width=7.5cm]{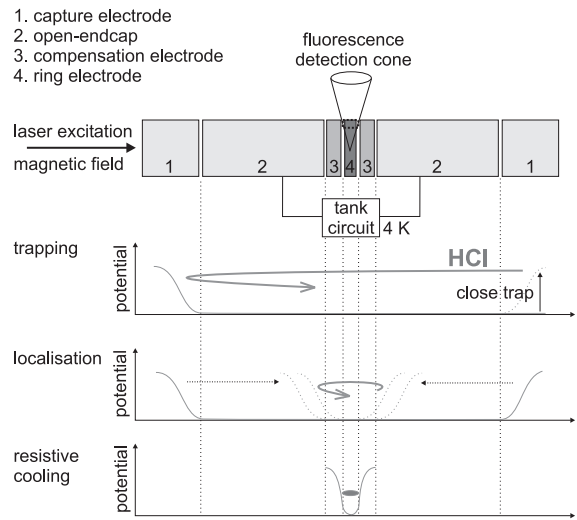}
\caption{Schematic of the proposed spectroscopy trap and the trap loading scheme.}
\label{fig3}
\end{center}
\end{figure}

The principle of resistive cooling is fairly straightforward and well-known: the axial motion of the HCI in the trap induces image charges in the open-endcap electrodes, which can readily be detected by a frequency resonant electronic tank circuit (LCR-filter) attached to the electrodes. If the tank circuit is at cryogenic temperatures (4~K), kinetic energy of the HCI is dissipated by the circuit, thus effectively cooling the ions. Since the cooling time varies with the charge state $q$ of the ions as $1/q^2$, this cooling scheme is very effective for HCI \cite{sta04}.

After the cooling procedure, the HCI will be focussed into a smaller cloud using the rotating wall technique \cite{dub99,hua98}. If the coupling of an ion plasma in a Penning trap is strong enough, the density of the ion cloud depends on its global rotation frequency. By applying RF voltages to a segmented ring electrode, the radial electric quadrupole field will rotate and drive the ion cloud. There are two limits for the rotation frequency $\nu$. The lower limit, which leads to the minimal density, is reached when $\nu$ equals the single-ion magnetron frequency $\nu_m$. At the upper limit, when $\nu$ is set to half the cyclotron frequency $\nu_c$, the density is maximum (Brillouin limit). (See {\it e.g.} \cite{win83,tho93} for the equations of motion of an ion in a Penning trap.)

Realistic values for the Penning trap are: a magnetic field of 6~T, an applied potential of 500~V, an ion cloud temperature of 4~K, and a trap parameter $d=17$~mm (see Ref \cite{gab89}). These values lead to the trap frequencies $\nu_m=23$~kHz, $\nu_c=36$~MHz, and an axial frequency $\nu_z=1.3$~MHz. If we set the rotation frequency to $\nu = 1$~MHz, the number density of a cloud of $10^5$ $^{207}$Pb$^{81+}$ ions is $5 \cdot 10^7$~cm$^{-3}$ (the Brillouin limit is $4.6 \cdot 10^8$~cm$^{-3}$). Such an ion cloud has a length of 6.8~mm and a diameter of 0.75~mm, which leads to an aspect ratio of 9.0. The Debye length of this cloud is 250~nm. The dependence of the ion cloud parameters on $\nu$ is shown in figure 4.

\begin{figure}[!t]
\begin{center}
\centering
\includegraphics[width=7.5cm]{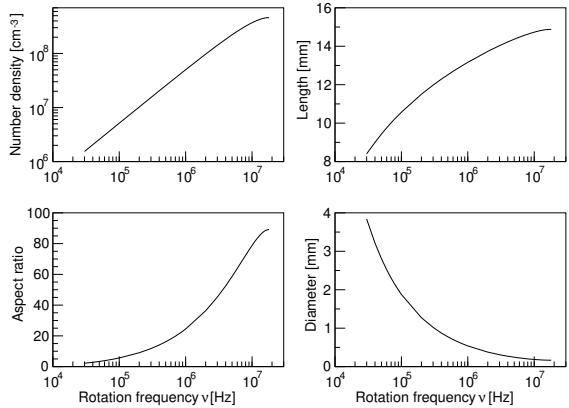}
\caption{The number density, length, aspect ratio and diameter of an ion cloud containing $10^5$ $^{207}$Pb$^{81+}$ ions, plotted as a function of the rotation frequency $\nu$. ($d=37.5$~mm)}
\label{fig4}
\end{center}
\end{figure}

A spectrum of the ground state hyperfine transition can be obtained by scanning the laser wavelength across the resonance while recording the fluorescence from the trapped decaying HCI. Repeated measurements on a single cloud of ions can be made, as detection is not destructive. To obtain the highest spectral resolution, the laser beam must be parallel to the magnetic field (trap axis), so as to avoid the large Doppler shift associated with the global rotation of the ion cloud. Laser excitation along the trap axis, on an ion cloud smaller than the laser beam, will also help to ensure that the transition is fully saturated.

\section{Calculated expected signal rates}
An ion temperature of 4~K corresponds to a Doppler width of 30~MHz. For $^{207}$Pb$^{81+}$ ions, the wavelength of 1020~nm corresponds to an upper state lifetime of 50~ms and a linewidth of 3~Hz. The laser power to fully saturate the Doppler-broadened transition is roughly 3~mW and is easily reached by {\it e.g.} a Ti:Sapphire or Ar$^+$ pumped dye laser. If the experiment is run in a continuous mode, assuming an overall detection efficiency of $4 \cdot 10^{-3}$, we expect to detect about $4 \cdot 10^3$ photon emissions per second for a completely saturated ion cloud. It should be possible to reduce the background signal to less than $10^2$ counts per second, yielding a typical S/N ratio of 40. Alternatively, if the laser excitation is pulsed with a duty cycle of 200~ms ($=4 \tau$), the signal is about $10^3$ counts per second, without any background from scattered laser light. These values are high enough to allow easy detection and measurement of the transition wavelength. Once the signal is seen, this will allow a wavelength determination to an accuracy which far exceeds the theoretical uncertainties. The linewidth of {\it e.g.} a Ti:Sapphire laser is 1~MHz at a wavelength of 1~$\mu$m, which is much smaller than the Doppler broadening of 30~MHz. Therefore the accuracy of the HFS measurement is estimated to be about $10^{-7}$, which is 3 orders of magnitude better than the current accuracy \cite{see98}.

\section{Acknowledgements}
This work is supported by the European Commission within the RTD programme FP5 (HPRI-CT-2001-50036 HITRAP). JRCP acknowledges the support by CONACyT, SEP and the ORS Awards Scheme.


\begin{thebibliography}{99}

\bibitem{bei00}
T. Beier, Phys. Rep. {\bf 339}, 79 (2000).

\bibitem{qui01}
W. Quint {\it et al.}, Hyp. Int. {\bf 132}, 457 (2001).

\bibitem{tdr03}
www.gsi.de/documents/DOC-2003-Dec-69-2.pdf

\bibitem{kla94}
I. Klaft {\it et al.}, Phys. Rev. Lett. {\bf 73}, 2425 (1994).

\bibitem{see98}
P. Seelig {\it et al.}, Phys. Rev. Lett. {\bf 81}, 4824 (1998).

\bibitem{cre96}
J.R. Crespo L\'opez-Urrutia, P. Beiersdorfer, D.W. Savin, K. Widmann, Phys. Rev. Lett. {\bf 77} 826 (1996).

\bibitem{cre98}
J.R. Crespo L\'opez-Urrutia, P. Beiersdorfer, K. Widmann, B.B. Birkett, A.-M. M\aa rtensson-Pendrill, M.G.H. Gustavsson, Phys. Rev. A {\bf 57}, 879 (1998).

\bibitem{bei01}
P. Beiersdorfer {\it et al.}, Phys. Rev. A {\bf 64}, 032506 (2001).

\bibitem{tho85}
R.C. Thompson, Rep. Prog. Phys. {\bf 48}, 531 (1985).

\bibitem{sha94}
V.M. Shabaev, J. Phys. B: At. Mol. Opt. Phys. {\bf 27}, 5825 (1994).

\bibitem{tom98}
M. Tomaselli, T. K\"uhl, P. Seelig, C. Holbrow, E. Kankeleit, Phys. Rev. C {\bf 58}, 1524 (1998).

\bibitem{sun98}
P. Sunnergren {\it et al.} Phys. Rev. A {\bf 58}, 1055 (1998).

\bibitem{zee96}
P. Zeeman, Versl. Kon. Ak. Wet. {\bf 5}, 181 (1896).

\bibitem{bre31}
G. Breit, I.I. Rabi, Phys. Rev. {\bf 38}, 2082 (1931).

\bibitem{gab89}
G. Gabrielse, L. Haarsma, S.L. Rolston, Int. J. Mass Spectr. Ion Proc. {\bf 88}, 319 (1989).

\bibitem{win75}
D.J. Wineland, H.G. Dehmelt, J. Appl. Phys. {\bf 46}, 919 (1975).

\bibitem{ver04}
J. Verd\'u, S. Djeki\'c, S. Stahl, T. Valenzuela, M. Vogel, G.Werth, T. Beier, H.-J. Kluge, W. Quint, Phys. Rev. Lett. {\bf 92}, 093002 (2004).

\bibitem{sta04}
S. Stahl (private communication).

\bibitem{dub99}
D.H.E. Dubin, T.M. O'Neil, Rev. Mod. Phys. {\bf 71}, 87 (1999).

\bibitem{hua98}
X.-P. Huang, J.J. Bollinger, T.B. Mitchell, W.M. Itano, Phys. Rev. Lett. {\bf 80}, 73 (1998).

\bibitem{win83}
D.J. Wineland, W.M. Itano, R.S. Van Dyck, Adv. At. Mol. Phys. {\bf 19}, 135 (1983).

\bibitem{tho93}
R.C. Thompson, Adv. At. Mol. Opt. Phys. {\bf 31}, 63 (1993).


\end{thebibliography}
\end{document}